\documentclass[12pt]{article}
\usepackage{amsfonts}
\usepackage{fancybox}
\usepackage{cite}
\usepackage[dvips]{graphicx}
\makeatletter
\@addtoreset{equation}{section}

\makeatother
\newcommand {\tr}{{\rm tr\,}}

\begin{document}
\begin{titlepage}
 
$\mbox{ }$
\begin{flushright}
\begin{tabular}{l}
\\
arXiv:0710.5873\\
TIFR/TH/07-29
\end{tabular}
\end{flushright}
~~\\
~~\\
~~\\

\vspace*{0cm}
    \begin{Large}
       \vspace{2cm}
       \begin{center}
        Monte Carlo Studies of the GWW Phase Transition\\
        in Large-$N$ Gauge Theories
       \end{center}
    \end{Large}

  \vspace{1cm}

\begin{center}
          Takehiro A{\sc zuma}\footnote{
e-mail address : azuma@theory.tifr.res.in},
          Pallab  B{\sc asu}\footnote{
e-mail address : pallab@physics.ubc.ca }$^{,\ast}$ and
          Spenta  R. W{\sc adia}\footnote{
e-mail address : wadia@theory.tifr.res.in\\ $\ast$ Address after Dec. 1, 2007: University of British Columbia, Vancouver, Canada.}
 \\

\vspace*{1cm}
{\it Department of Theoretical Physics, Tata Institute of Fundamental Research, Homi Bhabha Road, Mumbai, 400005, India}
\end{center}

\vfill

\begin{abstract}
\noindent
In the study of the small ten-dimensional Schwarzschild blackhole, the blackhole to string transition is an important problem.  In \cite{AlvarezGaume:2006jg}, a possible identification is made between the Gross-Witten-Wadia (GWW) type third-order large-$N$ phase transition in the boundary gauge theory and the string-black hole transition in the bulk. In this paper, we exhibit the existence of the GWW transition by Monte Carlo simulation in the zero mode bosonic action of the finite-temperature ${\cal N}=4$ SYM theory on $S^3$. Exhibiting  this transition in the truncated but highly non-trivial gauge theory implies that in the vicinity of the critical temperature $T_c$, the system goes critical, and the fluctuations give rise to universal formulas derived in \cite{AlvarezGaume:2006jg}. We also discuss the issue of $SO(6)$ R-symmetry breaking.

\end{abstract}
\vfill
\end{titlepage}
\vfil\eject

\tableofcontents

\section{Introduction}
Understanding the string-black hole transition is an important problem in string theory. The radius of a Schwarzschild blackhole becomes smaller with rising temperature and  at a certain temperature the blackhole transits to a gas of strings \cite{Susskind:1993ws,Horowitz:1996nw,Sen:1995in,Bowick:1986km,Bowick:1985af}. This is a difficult problem to address as it needs a proper understanding of non-perturbative effects in string theory. The finite temperature of a Schwarzschild blackhole breaks supersymmetry and string loop corrections are uncontrolled. In \cite{AlvarezGaume:2006jg} it was demonstrated that the problem of the string-blackhole transition can be formulated in a space-time with $AdS_5\times S^5$ boundary conditions. This enabled one to use the AdS/CFT correspondence \cite{9711200} to map the string-blackhole transition phenomenon to a Gross-Witten-Wadia (GWW) type phase transition \cite{Gross:1980he,Wadia:1979vk,Wadia:1980cp} in the boundary gauge theory defined on $S^3 \times R$. \footnote {One should caution against too literal an interpretation of the gauge theory result since at the cross over, the winding Polyakov line is non-zero, signaling that a black hole (without a space-time description, still persists at the phase transition and passes over into a stringy description only beyond the temperature at which the phase transition occurs. We would like to thank Juan Maldacena for a discussion of this point.}

The compactness and positive curvature of the space $S^3$ permits one to integrate out all other modes to get an effective multi-trace unitary matrix model for the zero mode of the Polyakov line. Based on the works \cite{9908001,Polyakov:2001af, Aharony:2003sx, Liu:2004vy,Spradlin:2004pp,Hallin:1998km,0502227,0506203}, this type of effective unitary matrix model was analyzed in  \cite{AlvarezGaume:2006jg} to show the existence of the GWW type transition. The $o(1)$ part of the gauge theory effective action was also calculated in a double scaled region near the transition temperature. The $o(1)$ part is universally given in terms of $F(t)$, where $F$ satisfies the following differential equation 
\begin{equation}
{\partial_t}^2{F(t)} = -f^2(t). \label{painleve}
\end{equation}
and $f(t)$ is the Painleve II function, and $t$ is a scaled variable proportional to $(T-T_c)N^{\frac{2}{3}}$.

The derivation of the effective unitary matrix model from the gauge theory is a subtle one. In a weakly coupled gauge theory one may demonstrate this explicitly in perturbation theory at large $N$ \cite{Aharony:2005bq}. However the situation is less clear in the strong coupling regime. One difficulty comes from the Gregory-Laflamme transition for a small $AdS_5\times S^5$ blackhole. This transition breaks the $SO(6)$ symmetry of $S^5$ and the question arises whether there is a new zero mode associated with this transition, and whether the unitary matrix model is a good description after this transition. 
In  \cite{AlvarezGaume:2006jg} it has been shown, using a supergravity analysis within the AdS/CFT correspondence, that even at strong coupling, the unitary matrix model serves as an effective description. 

Given the physical relevance of the GWW transition, it is important to see if this phenomenon occurs when one is not dealing with the dynamics (however complicated) of a single unitary matrix or the quantum mechanics of a single unitary matrix. It is not at all obvious that this large-$N$ transition occurs in more complicated models of non-commuting matrices and gauge theories. In the past this question has been explored by Douglas and Kazakov\cite{Douglas:1993iia} in their study of two-dimensional Yang-Mills theory on $S^2$. However this problem too, gets recast into a problem of a single unitary matrix because the partition function turns out to be the heat kernel on the unitary group.

In order to answer these questions, there seems to be no analytic tools as is usually the case with complicated dynamical problems. Hence we use numerical Monte Carlo methods to explore and exhibit the large-$N$ GWW transition and also study the question of R-symmetry breaking at large $N$. Since the full ${\cal N}=4$ SYM theory on $S^3 \times R$ is too difficult, in the first run we study the gauge theory restricted to the zero modes of the bosonic sector. It is likely that this reduction captures the essential features of the dynamics. It is motivated by the fact that the metric of the small Schwarzschild black hole is uniform on $S^3$. Regarding fluctuations in the bulk, the zero mode gauge theory has correspondence with fluctuations in the bulk which are independent of $S^3$ and only depend on the radial $AdS_5$ coordinate and time.

The importance of exhibiting  this transition lies in the fact that in the vicinity of the critical temperature $T_c$, the system goes critical and the fluctuations give rise to universal formulas (\ref{painleve}) which solely depend on the multi-critical point which is characterized by the exponent ${\frac{2}{3}}$ in the scaling law $(T-T_c)\sim N^{-{\frac{2}{3}}}$. Hence the formulas for the black hole cross over which were derived using the effective unitary matrix model in \cite{AlvarezGaume:2006jg} are also valid while working directly with the zero mode sector of the gauge theory.

This paper is organized as follows. In Section 2, we introduce the zero mode action of the bosonic part of the ${\cal N}=4$ SYM theory on $S^3 \times R$.  In Section 3, we discuss the numerical studies of the GWW-type phase transition. In Section 4, we study the $SO(6)$ R-symmetry using Monte Carlo simulation. Section 5 is devoted to conclusions and future outlook. 
 
\section{The model: ${\cal N}=4$ SYM theory reduced on $S^3$}
We study the ${\cal N}=4$ SYM theory, when all the bosonic fields are restricted to their zero modes on $S^3$. 
\begin{eqnarray}
 & & Z = \int dM dA e^{-S'}, \textrm{ where } \\ 
& & S' = N \int^{\beta}_{0} dt \left( \tr \sum_{\mu=1}^{D} (D_{t} M_{\mu}(t))^{2} - \frac{\lambda}{2} \tr \sum_{\mu,\nu=1}^{D} [M_{\mu}(t), M_{\nu}(t)]^{2} + m^{2} \sum_{\mu=1}^{D} \tr M^{2}_{\mu} (t)  \right), \nonumber \\ \label{action}
\end{eqnarray}
and $D_{t}$ is a covariant derivative defined by 
\begin{eqnarray}
 D_{t} M_{\mu}(t) = \partial_{t} M_{\mu}(t) - i [A(t), M_{\mu}(t)]. \label{covariant}
\end{eqnarray}
$D$ is the dimensionality of the model, and the dynamical variables $A(t)$ and $M_{\mu}(t)$ ($\mu= 1,2, \cdots, D$) are $N \times N$ Hermitian matrices, which can be regarded as the gauge field and the $SO(D)$ adjoint scalars, respectively. This model has a $U(N)$ gauge symmetry
\begin{eqnarray}
 M_{\mu}(t) \to g(t) M_{\mu}(t) g^{\dagger}(t), \hspace{3mm} A(t) \to g(t) A(t) g^{\dagger}(t) + ig(t) \frac{d g^{\dagger}(t)}{dt}. \label{gauge-tr}
\end{eqnarray}
The Euclidean time $t$ in the action (\ref{action}) has a finite extent $\beta$, which is the inverse temperature $\beta = 1/T$. Both the gauge and the scalar fields obey the periodic boundary conditions
\begin{eqnarray}
 A(t + \beta) = A(t), \hspace{3mm} M_{\mu}(t+ \beta) = M_{\mu}(t). \label{periodic_bc}
\end{eqnarray}
While this model has three parameters, $\beta$, $\lambda$ and $m$, these are not independent of each other, as $m$ can always be set to unity by the following redefinitions
\begin{eqnarray}
 \beta \to \frac{\beta}{m}, \hspace{3mm} \lambda \to \frac{\lambda}{m^{3}}  \label{reparametrization}
\end{eqnarray}
and rescaling of the fields
\begin{eqnarray}
 A(t) \to \frac{1}{m} A(t), \hspace{3mm} M_{\mu}(t) \to m^{\frac{1}{2}} M_{\mu}(t). \label{repara_field}
\end{eqnarray}
The periodic boundary condition (\ref{periodic_bc}) prevents us from fixing the $A=0$ gauge. However we can fix a gauge where the gauge field is static and diagonal:
\begin{eqnarray}
 A = \frac{1}{\beta} {\rm diag} (\alpha_{1}, \alpha_{2}, \cdots, \alpha_{N}), \label{gauge_fixing}
\end{eqnarray}
where $\alpha_a \in (- \pi, \pi]$. The indices $a,b, \cdots$ run over $1,2,\cdots, N$. This gives rise to the Fadeev-Popov term
\begin{eqnarray}
 S_{{\rm f.p.}} =  - \sum_{a,b=1, a \neq b}^{N} \log \sin | (\alpha_a - \alpha_b)/2 |, \label{fp_ghost}
\end{eqnarray}
whose derivation is given in full detail in  \cite{0310286,0601170}. In the following, we study the action 
\begin{eqnarray}
 S = S' + S_{{\rm f.p.}}. \label{action_gf}
\end{eqnarray}
We study the model (\ref{action_gf}) numerically by Monte Carlo simulation. The details of the algorithm are given in \cite{07063517}\footnote{In \cite{07061647,07074454}, they discuss the simulation of the supersymmetric gauge theory at finite temperature using Fourier transformation, instead of lattice discretization.}. We simulate the model with the time direction discretized. We apply the heat bath algorithm to the scalar fields and Metropolis algorithm to the gauge field, respectively. It turns out that taking 10 lattice points of the time direction is enough and that increasing the lattice points further does not affect the result. 

\section{GWW phase transition} 
In this section, we study the GWW type phase transition of the simplest unitary matrix model, for which an analytical solution is available \cite{Gross:1980he,Wadia:1979vk,Wadia:1980cp}. This model is useful to test the accuracy of the numerical method.

\subsection{The $trU^{\dagger} + trU$ model}
We start with the numerical simulation of the unitary matrix model consisting only of $\tr U$ without adjoint scalar fields, where 
\begin{eqnarray}
 U = {\cal P} \exp \left( i \int^{\beta}_{0} dt A(t) \right). \label{polyakov} 
\end{eqnarray}
${\cal P}$ denotes the path-ordered product. We consider the partition function
\begin{eqnarray}
 Z_{g} = \int dU \exp \left( \frac{Ng}{2} (\tr U +\tr U^{\dagger} )\right). \label{pf_0502227}
\end{eqnarray}
and define $u_n = \frac{1}{N} \tr U^n$ for an integer $n$. In the static and diagonal gauge (\ref{gauge_fixing}), 
\begin{eqnarray}
u_n = \frac{1}{N} \sum_{a=1}^{N} e^{i n \alpha_a }. 
\end{eqnarray}
These are the moments of the density of eigenvalues: $u(\alpha)=\frac{1}{2\pi}\sum_{a=1}^{N}\delta (\alpha-\alpha_a)$.

The first two moments are given by 
\begin{eqnarray}
 & & \langle |u_1| \rangle =\left\{ \begin{array}{ll} \frac{g}{2} & (\textrm{$g<1$}) \\ 1 - \frac{1}{2g} & (g>1)  \end{array} \right. \nonumber \\
 & &  \langle |u_2| \rangle =\left\{ \begin{array}{ll} 0 & (\textrm{$g<1$}) \\ 1 - \frac{2}{g}+\frac{1}{g^2} & (g>1).  \end{array} \right. \label{tru_analytic}
\end{eqnarray} 

The third-order transition at the point $g=1$ is the GWW transition. This is a transition between the gapped and ungapped phases of the eigenvalue distribution of the unitary matrix model.  For a generic unitary matrix model, all $u_n$'s  show a similar non-analytical behavior like $u_1$, because near the gap opening point, the relevant operator is given by a linear combination of $u_n$\cite{AlvarezGaume:2006jg}.

\begin{figure}[htbp]
   \begin{center}
    \scalebox{0.65}{\includegraphics{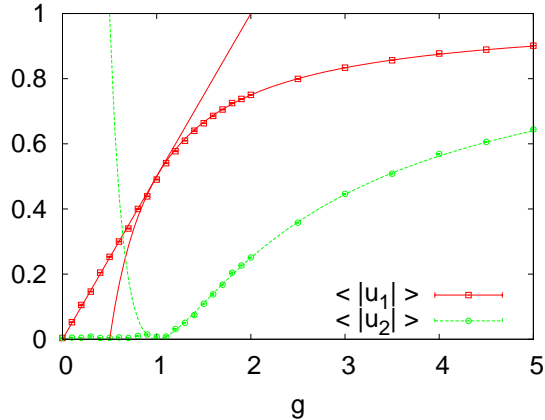}}
    \end{center}
  \caption{The vacuum expectation values $\langle |u_{1,2}| \rangle$ against $g$ for $N=128$.}
     \label{gw-0502227}
  \end{figure}

We first verify this result numerically using Monte Carlo simulation. To this end, we take static and diagonal gauge (\ref{gauge_fixing}) and add the Fadeev-Popov term (\ref{fp_ghost}). Namely, we apply the Metropolis algorithm to the action
\begin{eqnarray}
  - \frac{Ng}{2} (\tr U +\tr U^{\dagger}) - \sum_{a,b=1, a \neq b}^{N} \log \sin |\frac{\alpha_a - \alpha_b}{2}|.
\end{eqnarray}
We plot the VEV's $\langle |u_{1,2}| \rangle$ against $g$ in figure \ref{gw-0502227} for $N=128$, and find that they actually agree with the result (\ref{tru_analytic}). 
 
\subsection{GWW phase transition in the gauge theory reduced on $S^3$}
We next study the saddle point of the gauge field by adding the chemical potential $\mu (\tr U + \tr U^{\dagger})$ to the action (\ref{action_gf}). Namely, we study the matrix model
\begin{eqnarray}
 S_{g} = S' + S_{\rm g.f.} + N \beta \mu  (\tr U + \tr U^{\dagger}), \label{6d_chemical}
\end{eqnarray}
where the terms $S'$ and $S_{\rm g.f.}$ are defined in (\ref{action}) and (\ref{fp_ghost}), respectively, and $U$ is the Polyakov line defined in $(\ref{polyakov})$.

\subsubsection{$D=2$ case}
\begin{figure}[htbp]
   \begin{center}
    \scalebox{0.65}{\includegraphics{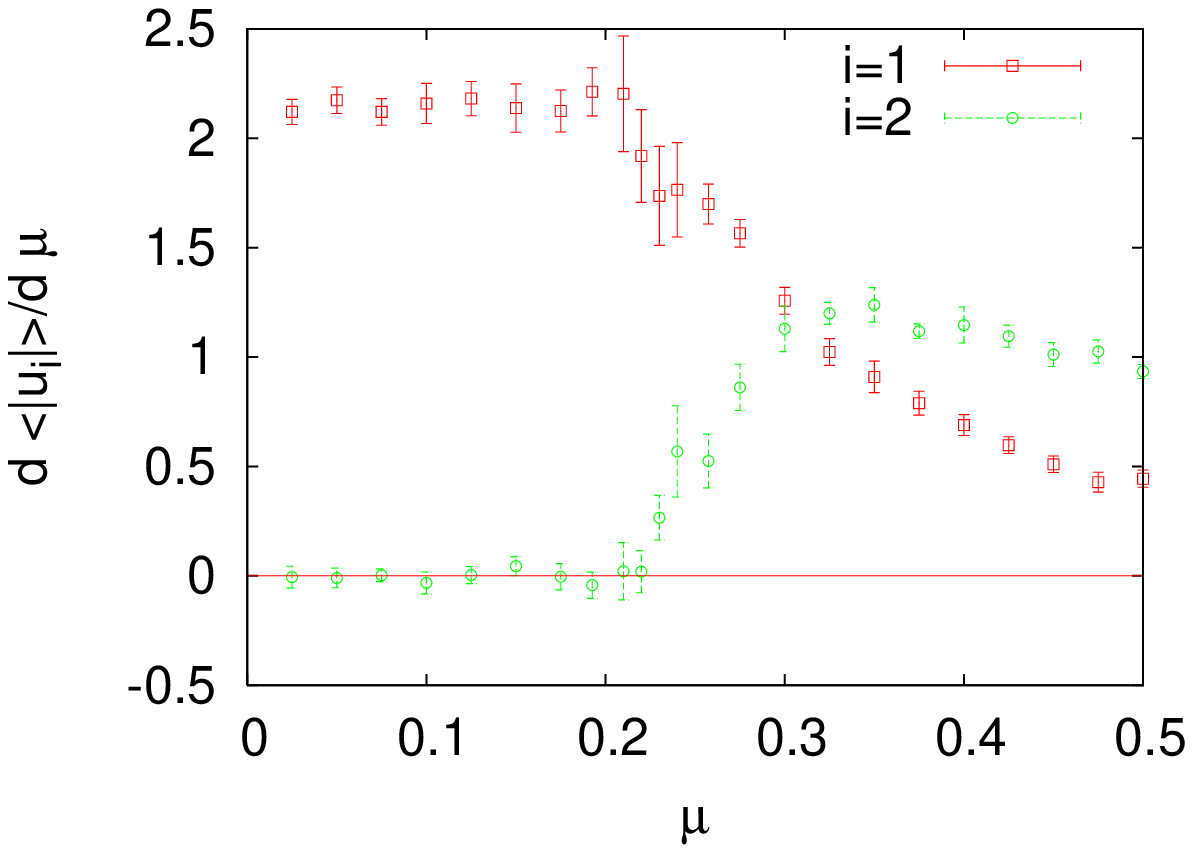}
                   \includegraphics{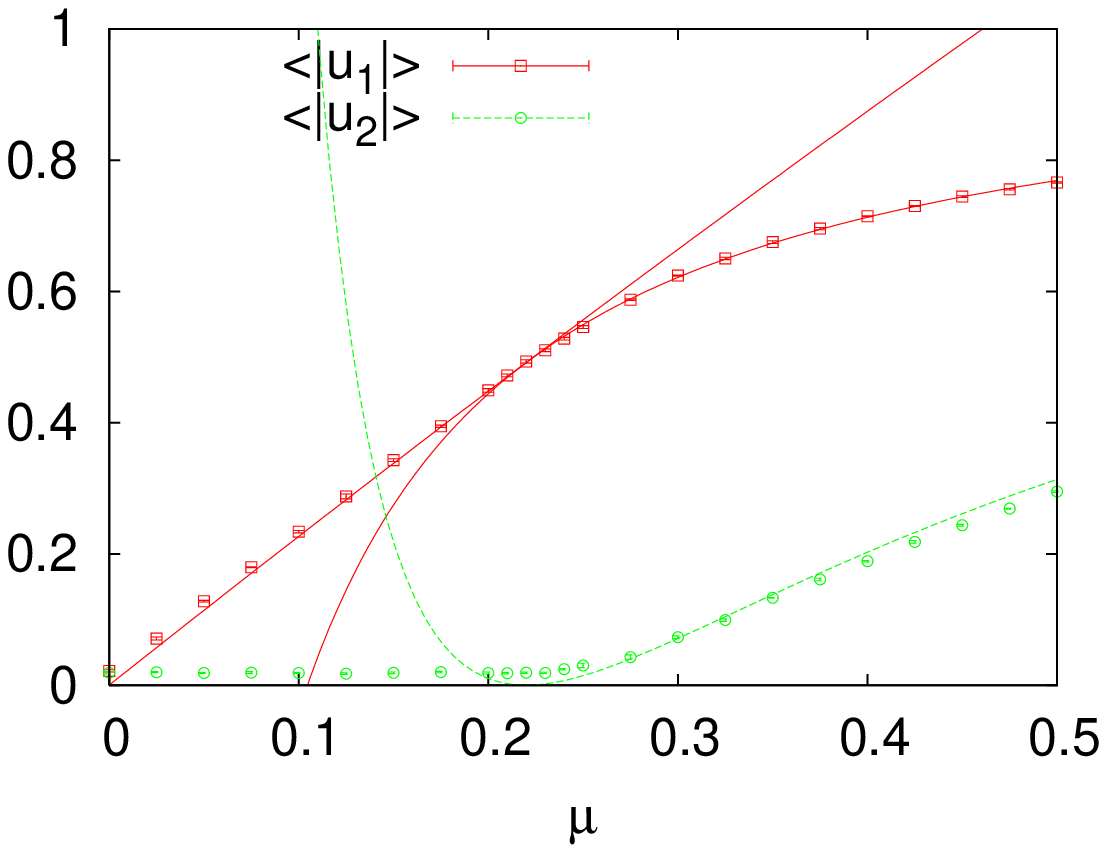}}
    \end{center}
  \caption{The vacuum expectation values $\frac{d \langle |u_{1,2}| \rangle}{d \mu}$ (left) and $\langle |u_{1,2}| \rangle$ (right) against $\mu$ for $D=2$, $\lambda =m =1.0$ and $N=48$.}
\label{d2-case}
  \end{figure}

We first study the $D=2$ case, in which the numerical simulation of large $N$ is reachable at a reasonable CPU time. The phase transition of the one-dimensional matrix quantum mechanics with respect to the temperature has been studied in  \cite{0406210,0508077,07043183,07063517} in the absence of the chemical potential. The Polyakov line $\langle |u_1| \rangle$ is small in the low-temperature region, while it is large in the high-temperature region. 
We focus on the low-temperature region $\beta = 2.0$, in which the Polyakov line $\langle |u_1| \rangle$ is small for $\mu=0.0$. We plot the result of the $D=2$, $\lambda =m =1.0$ and $N=48$ case in figure \ref{d2-case}.  
  
The graph above indicates a signature of the phase transition (possibly third or higher order) near the critical point 
\begin{eqnarray}
\mu_c \simeq 0.22, \label{crit-num}
\end{eqnarray}
at which the Polyakov line is $\langle |u_1| \rangle=0.5$. This 
is expected from the fact that $\langle |u_2| \rangle \ll \langle |u_1| \rangle$ near the transition point. To understand the nature of the transition we first numerically plot the derivative $\frac{d \langle |u_1| \rangle}{d\mu}$ in figure \ref{d2-case} (left).
The derivative seems to be continuous, and hence the possible transition should at least be of third order. Numerical errors prevent us from going further and calculating the higher derivatives directly from our data. Instead, in figure \ref{d2-case} (right) we try to fit our data with analytic functions in the regime $\mu <\mu_c$ and $\mu > \mu_c$ and extrapolate the information about derivatives from the fitted functions. It should be noted that the fitted functions do not necessarily represent the correct analytic form of the exact answer, but they can be viewed as a close approximation.  

We fit the VEV $\langle |u_1| \rangle$ with the function
\begin{eqnarray}
 \langle |u_1| \rangle = \left\{ \begin{array}{ll}  q_1 \frac{\mu}{\mu_c} + r_1 (\frac{\mu}{\mu_c})^{2}, & (\mu < \mu_c), \\ 1 - q_2 (\frac{\mu}{\mu_c})^{-1} - r_2 (\frac{\mu}{\mu_c})^{-2}, & (\mu > \mu_c), \end{array} \right. \label{u1_fit}
\end{eqnarray}
We exploit the fact that in the large-$N$ limit, $\langle | u_1| \rangle$ is 0  at $\mu=0$ and  $\langle | u_1| \rangle \to 1$ as $\mu \to \infty$. And from the fact that $\langle | u_1 | \rangle$ and its first derivative $\frac{d \langle |u_1| \rangle}{d \mu}$ are continuous at the critical point $\mu =\mu_c$, we obtain the condition for $r_1$ and $r_2$.
\begin{eqnarray}
 r_1 = \frac{1}{2} (1 - \frac{3}{2} q_1 - \frac{1}{2} q_2), \hspace{3mm} r_2 = \frac{1}{2} (1 - \frac{1}{2} q_1 - \frac{3}{2} q_2). \label{r-s}
\end{eqnarray}
The parameters $q_{1,2}$ are fitted as
\begin{eqnarray}
 q_1 = 0.503542 \pm 0.01181, \hspace{3mm} q_2 = 0.53791 \pm 0.003644. \label{q-s}
\end{eqnarray}
In this case, the coefficients $r_{1,2}$ are $r_1 = -0.0121$ and $r_2 = -0.029318$, respectively. We find that the contribution of the terms $r_1 (\frac{\mu}{\mu_c})^{2}$ for $\mu < \mu_c$ and $r_2 (\frac{\mu}{\mu_c})^{-2}$ for $\mu > \mu_c$ is small compared to the rest of the terms in (\ref{u1_fit}). Since its second derivative $\frac{d^2 \langle |u_1| \rangle}{d \mu^2}$ is discontinuous at the critical point $\mu = \mu_c$, this system undergoes the GWW type third-order phase transition.

The VEV $\langle |u_2 | \rangle$ is small when $\mu < \mu_c$, and in this region $\langle |u_2 | \rangle$ is closer to zero at larger $N$. When $\mu > \mu_c$, it is fitted with the following function similarly to the unitary matrix model. 
\begin{eqnarray}
 \langle |u_2 | \rangle =1 - \frac{2 \mu_c}{\mu} + \frac{\mu_c^2}{\mu^2}, \hspace{3mm} (\mu > \mu_c). \label{u2_fit}
\end{eqnarray}

\begin{figure}[htbp]
   \begin{center}
    \scalebox{0.65}{\includegraphics{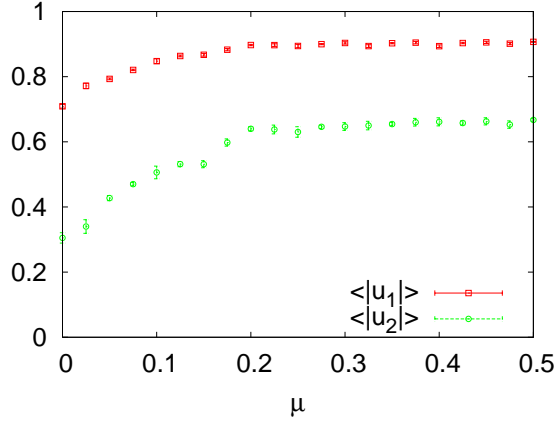}}
    \end{center}
  \caption{The vacuum expectation values $\langle |u_{1,2}| \rangle$ against $\mu$ in the $D=2$, $\beta=0.2$, $\lambda =m =1.0$ case for $N=48$.}
\label{d2-high}
  \end{figure}
Next, we turn our attention to the high-temperature case $\beta =0.2$, in which the VEV's of the Polyakov line $\langle |u_{1,2} | \rangle$ are large even in the absence of the chemical potential. We plot $\langle | u_{1,2} | \rangle$ against $\mu$ for $D=2$, $\beta=0.2$, $\lambda = m =1.0$ case for $N=48$ in figure \ref{d2-high}. In contrast to the low-temperature case, we find that there is no GWW type third-order phase transition in this case and that the VEV's of the Polyakov line increase monotonically.

\subsubsection{$D=6$ case} 
\begin{figure}[htbp]
   \begin{center}
    \scalebox{0.65}{\includegraphics{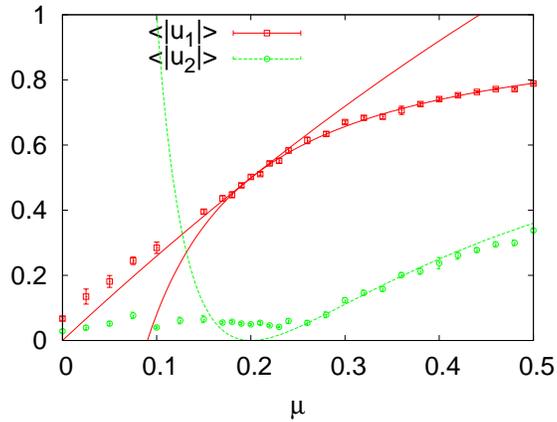}}
    \end{center}
  \caption{The vacuum expectation values $\langle |u_{1,2}| \rangle$ against $\mu$ in the $D=6$, $\beta=2.0$, $\lambda =m =1.0$ case for $N=16$.}
\label{d6-case}
  \end{figure}

We next study a different dimensionality, $D=6$. Similarly, we plot the VEV's $\langle |u_{1,2}| \rangle$ against $\mu$ in the $D=6$, $\beta=2.0$, $\lambda =m =1.0$ case for $N=16$ in figure \ref{d6-case}. We read off the critical point as
\begin{eqnarray}
 \mu_c \simeq 0.20.
\end{eqnarray}
Then, we fit them with the functions (\ref{u1_fit}) and (\ref{u2_fit}). In this case, the parameters are 
\begin{eqnarray}
 q_1 = 0.53803 \pm 0.02015, \hspace{3mm} q_2 = 0.542291 \pm 0.006888. \label{q-s-d6}
\end{eqnarray}
The coefficients $r_{1,2}$ are $r_1 = -0.039095$ and $r_2 = -0.0412257$, which suggests that the contribution of the $r_{1,2}$ terms is smaller than that of the rest of the terms in (\ref{u1_fit}). We find that the result is similar to the $D=2$ case.

\section{$SO(6)$ R-symmetry breaking}
In this section, we study the spontaneous breaking of the $SO(6)$ R-symmetry of the model (\ref{action_gf}) by Monte Carlo simulation. Throughout this section, we focus on the $D=6$ dimensional case. In analogy with the IKKT-type matrix model \cite{9612115}, we consider the following observable \cite{9811220,0003208,0005147,0104260,0108041,0108070,0412194},
\begin{eqnarray}
   I'_{\mu \nu} (t) = \frac{1}{N} \tr (M_{\mu}(t) M_{\nu}(t)).
\end{eqnarray}
In our case, we integrate the operator $I'_{\mu \nu} (t)$ with respect to the time direction and obtain the "integrated moment of inertia tensor"
\begin{eqnarray}
I_{\mu \nu} = \frac{1}{N} \int^{\beta}_{0} dt \tr M_{\mu}(t) M_{\nu}(t). \label{so6-observable}
\end{eqnarray}
We define the eigenvalues of this $6 \times 6$ matrix $I_{\mu \nu}$, which are all real positive,  as $\lambda_{\mu}$ with the specific order
\begin{eqnarray}
\lambda_{1} \geq \lambda_{2} \geq \cdots \geq \lambda_{6}.
\end{eqnarray} 
We consider the following $SO(6)$ invariant quantity \cite{9811220},
\begin{equation}
J=\frac{1}{6} I_{\mu\nu} I_{\mu\nu} - (\frac{1}{6} I_{\mu\mu})^2.
\end{equation}
This quantity measures the variance of the eigenvalue distribution of $I_{\mu \nu}$. Using large-$N$ factorization we get
\begin{eqnarray}
\langle J \rangle &=& \langle \frac{1}{6}  I_{\mu\nu} I_{\mu\nu} - (\frac{1}{6} I_{\mu\mu})^2 \rangle \\               
                  &\approx& \frac{1}{6}\langle I_{\mu\nu} \rangle  \langle I_{\mu\nu} \rangle  - \frac{1}{6^2} \langle I_{\mu\mu}^2 \rangle.
\end{eqnarray}
Using the fact that the VEV of any $SO(6)$ two-tensor is proportional to $\delta_{\mu\nu}$, i.e. $\langle I_{\mu\nu} \rangle=\lambda \delta_{\mu\nu}$, we get $\langle J \rangle=0$. This relationship is not true in general and we expect $\langle J \rangle \neq 0$ at finite $N$. In the case when $\langle J \rangle$ is non-zero, the width of the eigenvalue distribution of $I_{\mu \nu}$ is non-zero. The above scenario implies that the dominant contribution of the path integral comes from the configurations for which the eigenvalues of $I_{\mu\nu}$ are not equal and consequently the $SO(6)$ symmetry is broken. Hence by plotting the VEV's of the eigenvalues of $I_{\mu\nu}$ and measuring the width of the distribution, we can figure out the possibility of $SO(6)$ symmetry breaking at large $N$.  
This leads us to evaluate the VEV's of these eigenvalues $\langle \lambda_{\mu} \rangle$ in the large-$N$ limit. After diagonalization, the residual $SO(6)$ transformations permute the eigenvalues $\lambda_{\mu}$. Hence an unbroken $SO(6)$ symmetry implies,
\begin{equation}
\langle \lambda_{1} \rangle = \langle \lambda_{2} \rangle = \cdots = \langle \lambda_{6} \rangle.
\end{equation}
whereas a broken $SO(6)$ symmetry implies that for some $ \mu>\nu ,\langle \lambda_{\mu} \rangle > \langle \lambda_{\nu} \rangle$.
To this end, we extrapolate the large-$N$ limit from the simulation of finite $N$. If the eigenvalues $\langle \lambda_{\mu} \rangle$ are all equal in the large-$N$ limit, this suggests that the $SO(6)$ symmetry is unbroken.

\subsection{Dynamical gauge field}
\begin{figure}[htbp]
   \begin{center}
    \scalebox{0.65}{\includegraphics{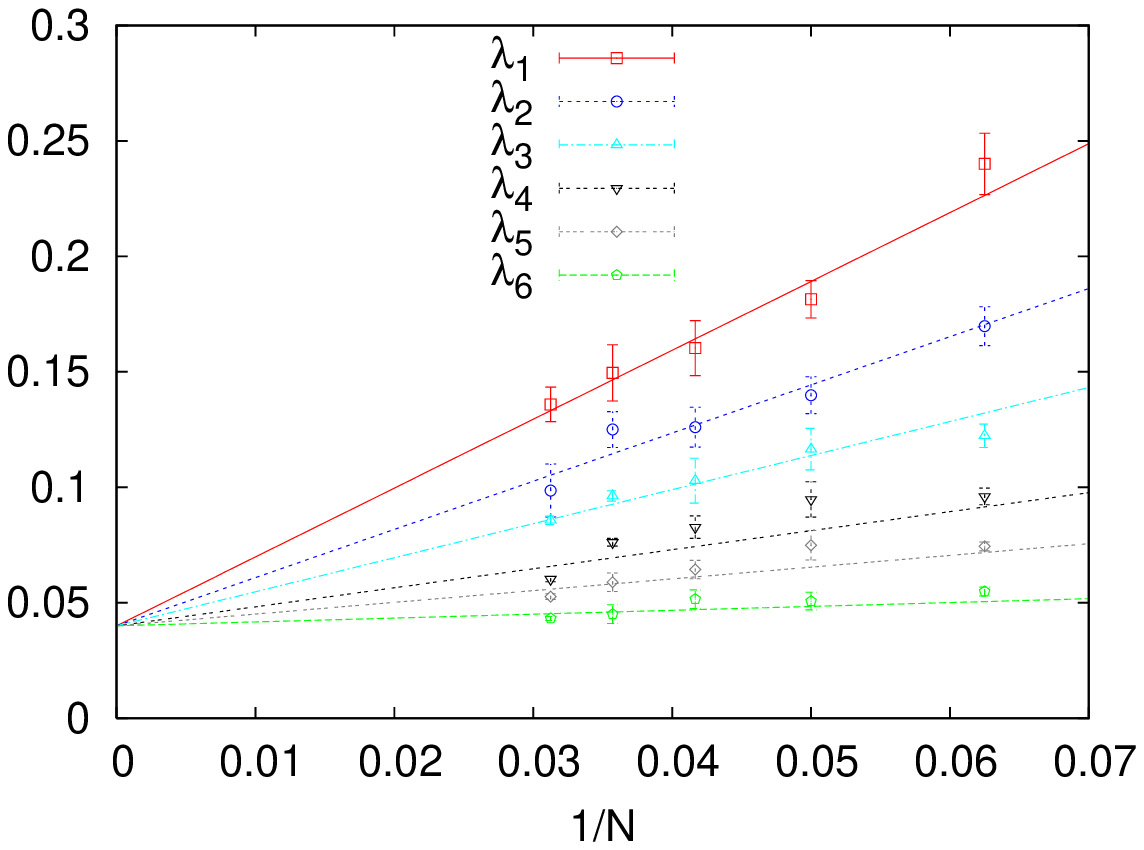}
                   \includegraphics{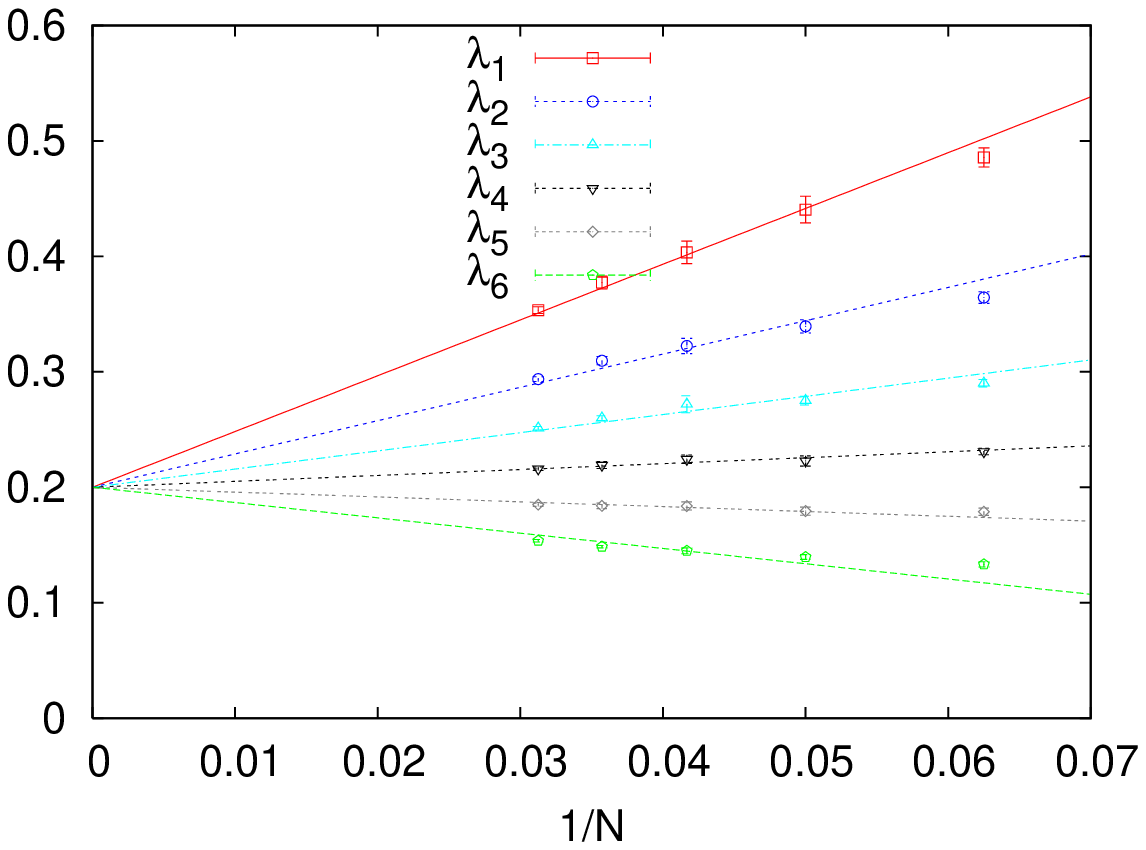}}
    \scalebox{0.65}{\includegraphics{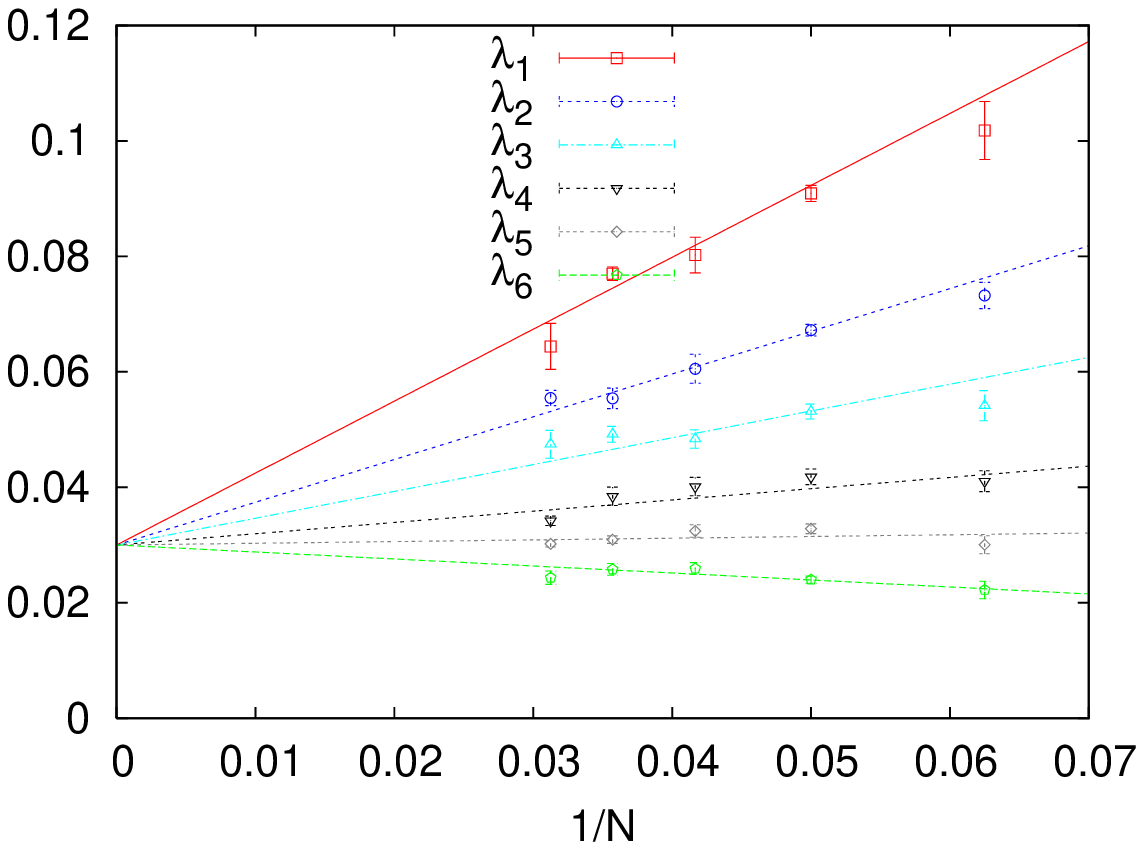}
                   \includegraphics{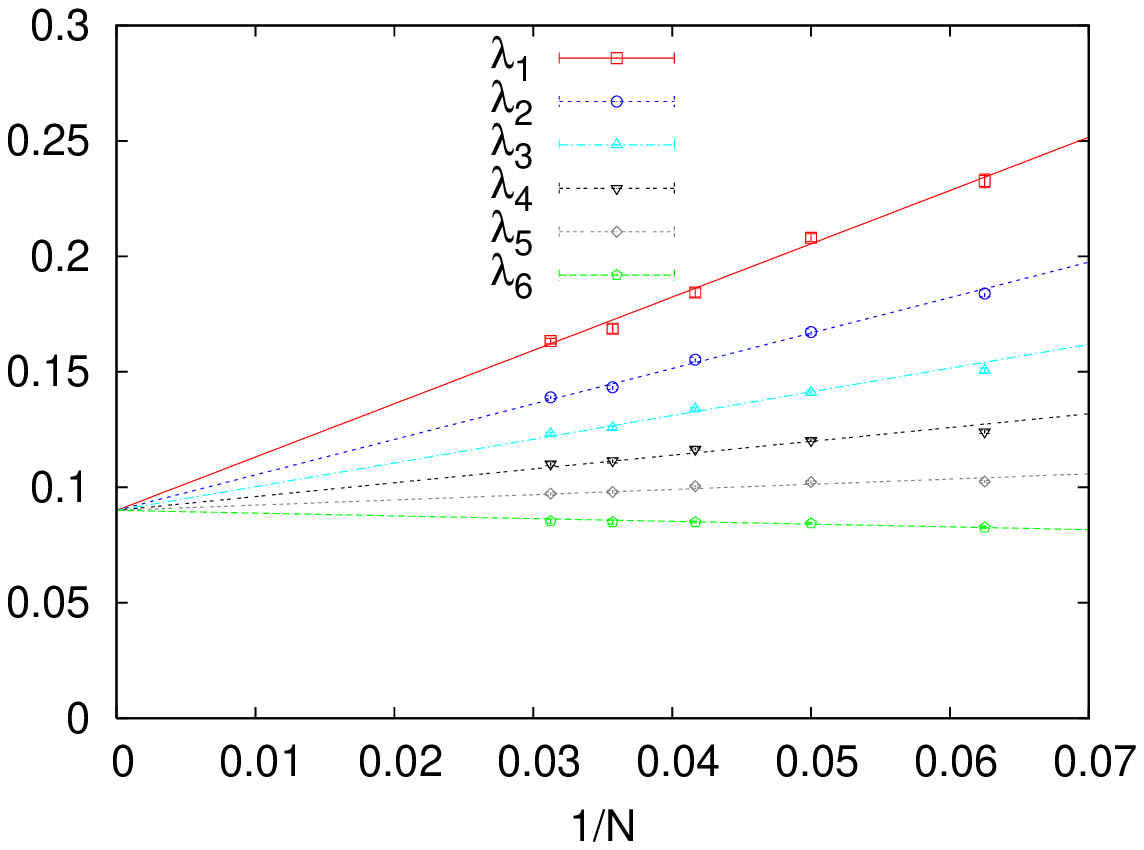}}
    \end{center}
  \caption{$\langle \lambda_{\mu} \rangle$ against $1/N$ for $m = 1.0$, and $N = 16, 20, 24, 28, 32$. $(\beta, \lambda)=(0.1, 0.1)$ (upper left), $(\beta, \lambda)=(1.0, 0.1)$ (upper right), $(\beta, \lambda)=(0.1, 1.0)$ (lower left), $(\beta, \lambda)=(1.0, 1.0)$ (lower right).}
     \label{teig1-vacuum}
  \end{figure}

We first study the $SO(6)$ R-symmetry breaking when gauge field $A$ is integrated. To this end, we update the gauge field $A$, as well as the scalar fields $M_{\mu}(t)$ via the usual algorithm.  We extrapolate the large-$N$ limit from the finite-$N$ results of $N=16, 20, 24, 28, 32$ for the high-temperature $\beta = 0.1$ and the middle-temperature $\beta =1.0$
 cases. We plot the eigenvalues $\langle \lambda_{\mu} \rangle$ against $1/N$ in figure \ref{teig1-vacuum}.
  
It turns out that the eigenvalues $\langle \lambda_{\mu} \rangle$ converge to the same value in the large-$N$ limit. This behavior is qualitatively the same for other parameter regions of the action (\ref{action_gf}). This indicates that the $SO(6)$ R-symmetry is unbroken in the matrix model (\ref{action_gf}).

\subsection{Uniform and Clumped configurations of the gauge field}
We next study the $SO(6)$ R-symmetry breaking in the specific configurations of the gauge field, which correspond to $AdS_5\times S^5$ and a black hole. To this end, we put the constraints on the gauge fields. In the following, we focus on the high-temperature $\beta=0.1$ and massive $m=1.0$ case, and take $\lambda=1.0$. 
\paragraph{1. Uniform distribution} \hspace{0mm} \\

\begin{figure}[htbp]
   \begin{center}
    \scalebox{0.6}{\includegraphics{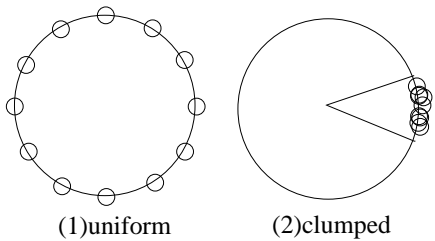}}
    \end{center}
  \caption{Distribution of the diagonal part of the gauge fields $\{ e^{i \beta \alpha_a} \}$ in (1)uniform and (2) clumped distribution.}
  \label{gauge-distribution}
  \end{figure}

We take the diagonal part of the gauge field (\ref{gauge_fixing}) as 
\begin{eqnarray}
\alpha_{a} = \frac{\pi}{N} (2a - N). \label{gauge-uni}
\end{eqnarray}
In this case, the Polyakov line $U$ satisfies $\langle |u_n| \rangle =0$ for any nonzero integer $n$. In the AdS/CFT correspondence uniform distribution, which is depicted in figure \ref{gauge-distribution} (1), corresponds to the $AdS_{5} \times S^5$ spacetime \cite{9803131,AlvarezGaume:2006jg}. To realize this configuration, we skip the Metropolis algorithm to update the gauge field $A$ and fix the configuration of the gauge field to be (\ref{gauge-uni}). We update only the scalar fields $M_{\mu} (t)$ via heat bath algorithm. 


\paragraph{2. Clumped distribution} \hspace{0mm} \\

In the clumped distribution, we constrain the gauge fields in a small region $\alpha_{a} \in [-\pi \epsilon, +\pi \epsilon]$, which is opposite to the gapped distribution. This distribution is depicted in figure \ref{gauge-distribution} (2). Similarly to the gapped distribution, we take $\epsilon = 0.05$. If $\alpha_a$ goes out of the region $[-\pi \epsilon, +\pi \epsilon]$, we automatically reject that configuration. This configuration coming from a gapped distribution of eigenvalues corresponds to the blackhole state as can be indicated by an analysis of large-$N$ perturbation theory around the gapped phase \cite{AlvarezGaume:2006jg}.

Similar to the case when we updated the configuration of the gauge field, we make a large-$N$ extrapolation of the eigenvalues $\langle \lambda_{\mu} \rangle$. 

\begin{figure}[htbp]
   \begin{center}
    \scalebox{0.65}{\includegraphics{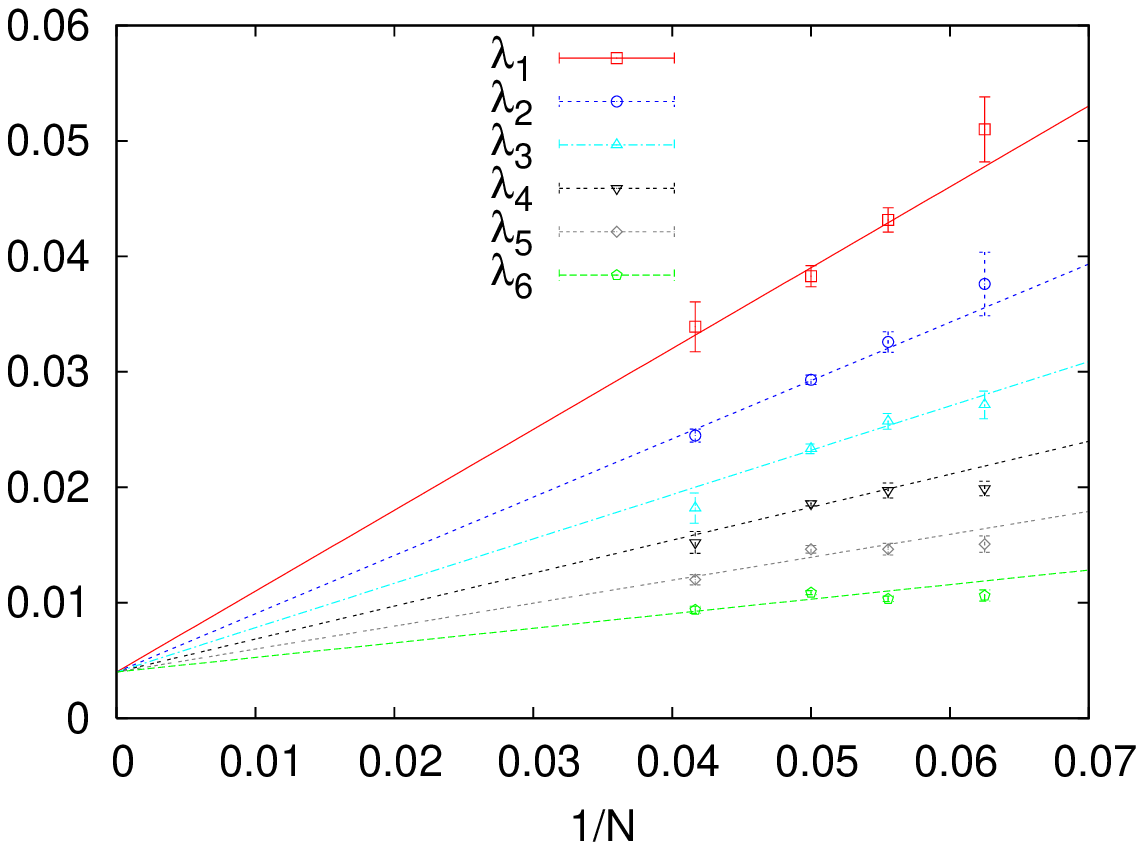}
                   \includegraphics{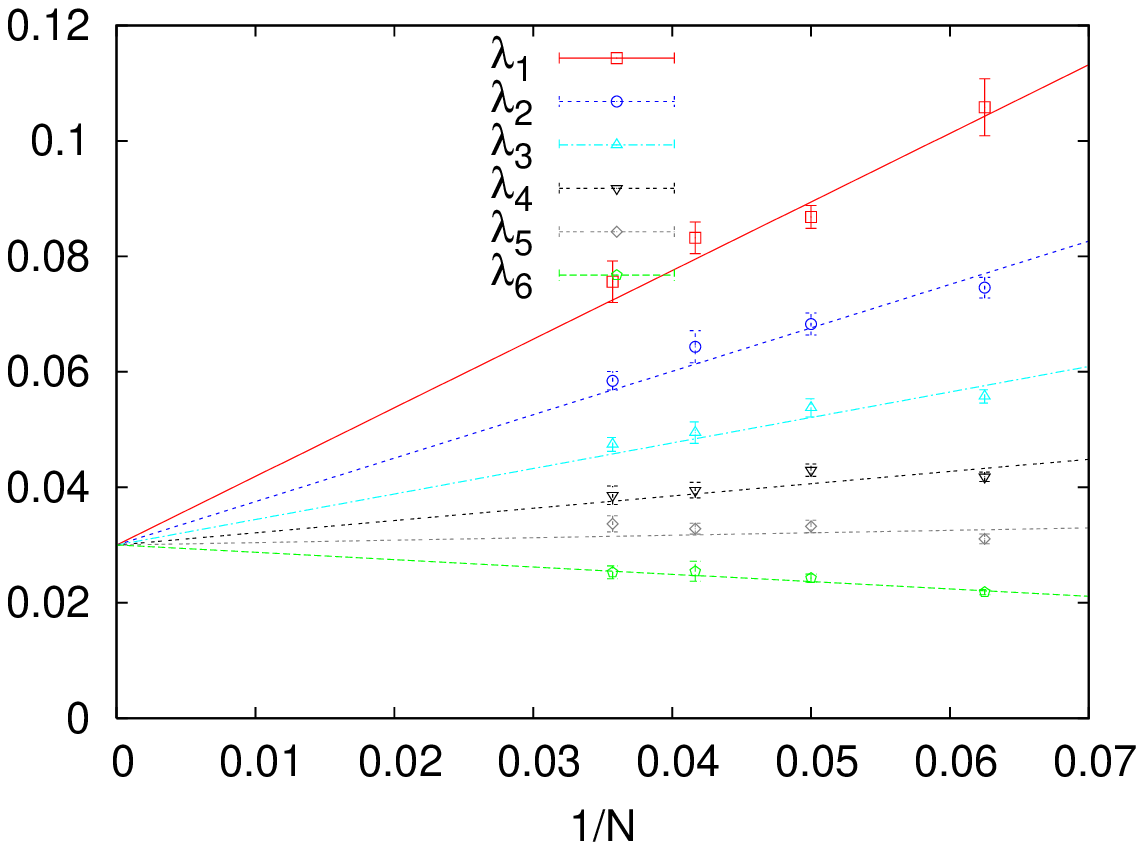}}
    \end{center}
  \caption{$\langle \lambda_{\mu} \rangle$ against $1/N$ for $\beta = 0.1$ and $\lambda = m =1.0$ in the uniform (left) and clumped (right) distribution.}
     \label{teig1-specific}
  \end{figure}

We plot in figure \ref{teig1-specific} the eigenvalues $\langle \lambda_{\mu} \rangle$ against $1/N$
 for the high-temperature $\beta=0.1$ and $\lambda =m =1.0$ case. In these cases, too, the eigenvalues $\langle \lambda_{\mu} \rangle$ converge to the same value at large $N$. We find that the $SO(6)$ R-symmetry of the scalar field is unbroken for these configurations of the gauge field.

\section{Conclusions}
In this paper, we have exhibited the GWW large-$N$ phase transition using Monte Carlo simulation in the zero mode reduction of the bosonic part of the ${\cal N}=4$ SYM theory on the $S^3 \times R$ space. We have studied the saddle point by adding a chemical potential to the reduced action, and observed a third-order phase transition in the large-$N$ limit. Its significance is that the large-$N$ transition signals critical behavior and the the properties of the model in the vicinity of the critical point are universal. Hence we expect that the $o(1)$ free energy is given by (\ref{painleve}). 
We have also numerically found that the $SO(6)$ R-symmetry is NOT spontaneously broken, in the large-$N$ limit.
In the $d=0$ and $d=1$ unitary matrix models the physical mechanism for the GWW transition is well understood. In the $d=0$ models the repulsion between eigenvalues, from the measure, and their attraction in the potential well, are competing effects which lead to this transition \cite{Gross:1980he,Wadia:1979vk}.\footnote{Recently Dutta and Gopakumar\cite{Dutta} have discussed the  multi-trace unitary matrix model in the large-$N$ limit, in terms of the saddle point in the space of the Young Tableaux density. This density and the eigenvalue density provide a very interesting phase space picture of the large-$N$ transition.} In the $d=1$ models the phase transition is signaled when the Fermi level reaches the hump (maximum) of the potential  \cite{Wadia:1980cp}. In the more complicated models we have explored, there are typically non-commuting matrices and explanation seems to be less obvious.

In the future it would be instructive to go beyond the zero mode approximation and develop numerical methods to include the variation of the fields on $S^3$.  Also, it would be interesting to be able to include the adjoint fermions of the gauge theory in the numerical calculation.




 \paragraph{Acknowledgment} \hspace{0mm} \\
 The authors would like to thank Rajesh Gopakumar, Shiraz Minwalla, Shingo Takeuchi and Toby Wiseman for valuable discussions. T.A. thanks Sourendu Gupta for help in the use of the computer system. Part of the simulations
were performed on the computer cluster of the Physics Theory group of the National Technical University of Athens. SRW would like to thank the J. C. Bose Fellowship of the Govt. of India.

\end{document}